\documentclass[reprint,prb,twocolumn,superscriptaddress]{revtex4-2} 
\usepackage{amsmath}
\usepackage{amsfonts}
\usepackage{graphicx}
\usepackage{color}
\usepackage{xcolor}
\usepackage{float}

\newcommand{\bs}{\boldsymbol}

\begin{document}
\title {Microwave detection of gliding Majorana zero modes in nanowires}
\author{Olesia Dmytruk}
\affiliation{JEIP, UAR 3573 CNRS, Coll\`ege de France, PSL Research University, F-75321 Paris, France}

\author{Mircea Trif}
\affiliation{International  Research  Centre  MagTop,  Institute  of  Physics,  Polish  Academy  of  Sciences, Aleja  Lotnikow  32/46,  PL-02668  Warsaw,  Poland}
\date{\today}

\begin{abstract}

We study a topological superconducting nanowire that hosts gliding Majorana zero modes in the presence of a microwave cavity field. We show that the cavity decay rate depends on both the parity encoded by the Majorana zero modes and their motion, in the absence of any direct overlap of their wave-functions. That is because the extended bulk states that overlap with both Majorana states, facilitate their momentum-resolved microwave spectroscopy, with the gliding acting as to modify the interference pattern via a momentum boost. Moreover, we demonstrate that these non-local effects are robust against moderate disorder in the chemical potential, and confirm the numerical calculations with an analytical low-energy model. Our approach offers an alternative to tunneling spectroscopy to probe non-local features associated with the Majorana zero modes in nanowires.

\end{abstract}

\maketitle

\section{Introduction} 

Topological phases of matter in the context of condensed matter physics emerged as a prominent research field following theoretical predictions of topological insulators and superconductors~\cite{hasan2010colloquium,qi2011topological}.
In particular, topological superconductivity is characterized by the emergence of the zero-energy bound states that
are equal superpositions of electrons and holes~--~Majorana zero modes (MZMs)~\cite{kitaev2001unpaired,beenakker2013search}. They obey non-Abelian statistics and are stable against disorder, making them potential building blocks for a topological quantum computer~\cite{kitaev2003fault,nayak2008non}. 

MZMs were predicted to emerge in a number of solid-state platforms including graphene-like systems~\cite{klinovaja2012electric,klinovaja2012helical,black2012edge,klinovaja2013giant,san2015majorana,kaladzhyan2017formation} and chains of magnetic adatoms~\cite{nadj2013proposal,klinovaja2013topological,braunecker2013interplay,vazifeh2013self}, with the most experimental efforts focused on semiconducting nanowires~\cite{lutchyn2010majorana,oreg2010helical}. 
Over the last years, many experimental works reported the observation of zero-bias conductance peaks~\cite{mourik2012signatures,deng2012anomalous,das2012zero,churchill2013superconductor,deng2016majorana,de2018electric} that were theoretically predicted to be a necessary signature of the presence of MZMs.
However, several theoretical works demonstrated that zero-bias conductance peaks in
experimental platforms consisting of a proximitized nanowire coupled to a 
quantum dot~\cite{deng2016majorana} could arise from the trivial Andreev bound states formed in the quantum dot~\cite{liu2017andreev,ptok2017controlling,reeg2018zero,hess2021local}.
In view of the controversy associated with zero-bias peak signatures of the MZMs, alternative ways to probe MZMs are highly needed at the moment. One of the approaches going beyond local transport measurements is based on  circuit quantum electrodynamics (QED)~\cite{blais2004cavity,wallraff2004strong}. It was theoretically predicted that MZMs in proximitized nanowires coupled to superconducting resonator could be probed via measuring cavity transmission coefficient~\cite{trif2012resonantly,cottet2013squeezing,dmytruk2015cavity,dmytruk2016josephson,dartiailh2017direct,cottet2017cavity,trif2018dynamic,trifPRL19,contamin2021hybrid}. However, these previous works that use quantum optics methods mainly focused on static MZMs, while their dynamics, which is crucial in view of using them as topological qubits, remains largely unexplored.

Here we fill this gap and investigate the dynamics of photons in a microwave cavity coupled to a topological superconducting nanowire in the ballistic regime that hosts gliding MZMs. We demonstrate that both the ground state parity encoded by the MZMs and their gliding dynamics influence the cavity field decay rate into the external lines that can be accessed experimentally. We determine that these effects originate from interference processes between the localised MZMs and the extended bulk states that are being ignited by the cavity, with the gliding acting as to modify the interference pattern via a momentum boost in the bulk states. This mechanism is similar to that in Refs.~\cite{AuslaenderScience02,TserkovnyakPRL02}, where  the momentum-resolved tunneling spectroscopy of a finite nanowire in the presence of a magnetic field (causing a momentum boost) reveals information about the confining potential landscape and the nature of excitations.  
\begin{figure}[t] 
\centering
\includegraphics[width=0.99\linewidth]{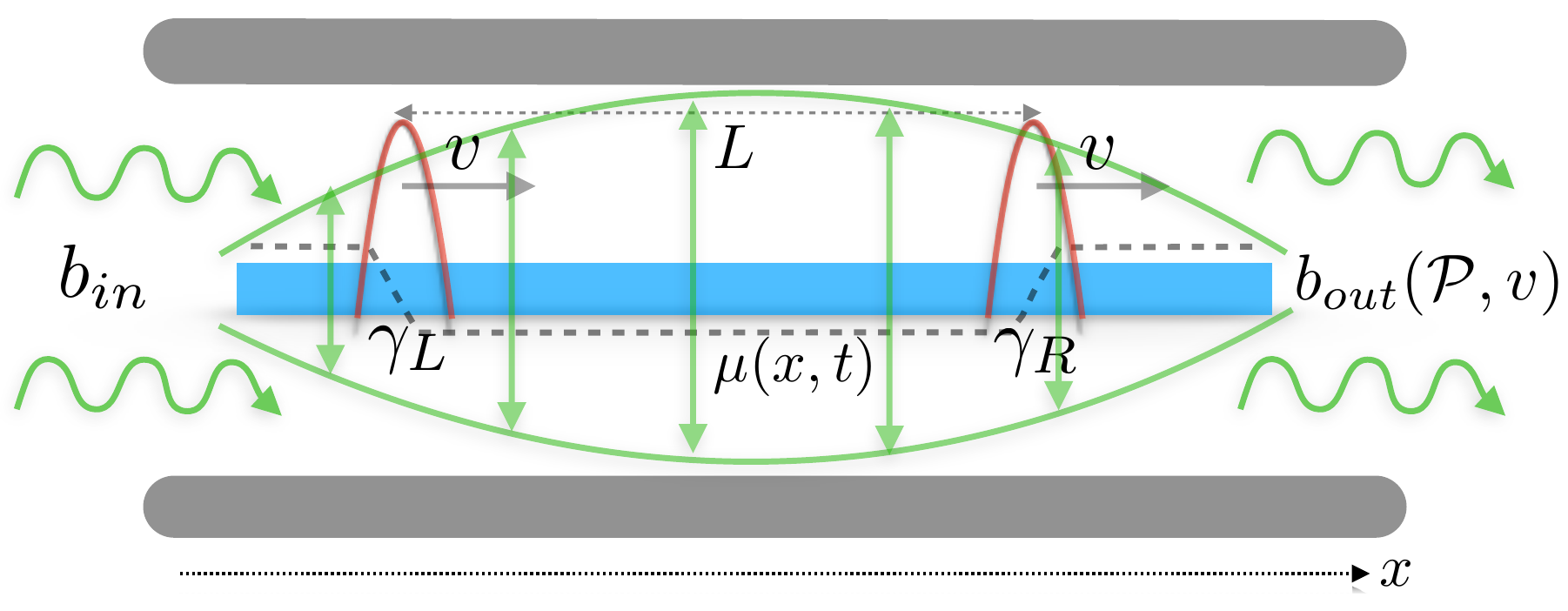}
\caption{Sketch of the topological superconducting nanowire-cavity system. The nanowire hosts two MZMs ($\gamma_{L,R}$) in the presence of an inhomogeneous and time-dependent chemical potential $\mu(x,t)\equiv\mu(x-vt)$. Both MZMs move with velocity $v$ and couple to the cavity field (in green). The photons transmitted outside the cavity, $b_{out}$, depend on both the fermionic parity $\mathcal{P}\equiv i\langle \gamma_L\gamma_R\rangle$ defined by the two MZMs and their velocity. As described in the text, this reflects the interplay of the non-locality of both the bulk states and the cavity field.}
\label{fig:Fig1}
\end{figure}

The paper is organised as follows: in Sec. II we introduce the model and determine the general form of the electronic susceptibility. Then, in Sec. III, we use a tight-biding approach to describe an experimentally relevant SO coupled nanowire hosting MZMs, both when they are static and gliding. In Sec. IV we introduce a low-energy model that  captures analytically the main features found in lattice numerics. Finally, we end up in Sec. V with conclusions. 


\section{Model and main results}

The Hamiltonian describing a one-dimensional topological superconductor (SC) in Fig.~\ref{fig:Fig1} is 
\begin{align}
    \mathcal{H}_{tot}(t)&=\frac{1}{2}\int dx\,\Psi^\dagger( x)[H_{el}(t)+H_{e-p}]\Psi(x)+H_{p}\,,
\end{align}
where $H_{el}(t)$ is the (time-dependent) Bogoliubov-de-Gennes (BdG) Hamiltonian describing the nanowire electrons, $H_{e-p}$ is the electron-photon coupling, while $H_{p}=\omega_ca^\dagger a$ describes the single-mode cavity photons with frequency $\omega_c$ and creation (annihilation) operator $a^\dagger$ ($a$). The electronic field operators can be written as $\Psi(x)=(\psi_\uparrow( x),\psi_\downarrow(x),\psi^\dagger_\downarrow(x),-\psi^\dagger_\uparrow(x))^T$, so that the electron-photon coupling Hamiltonian reads $H_{e-p}=g(x)\tau_z(a^\dagger+a)$, with $g(x)$ being the position-dependent coupling strength and $\tau_z$ is the Pauli matrix acting in the particle-hole subspace. Next, we aim to unravel the effect of the electrons on the photonic dynamics and, to keep the approach general, we will not make reference to the explicit form of $H_{el}(t)$ in this section. The equation of motion for the photonic operator up to second order in electron-photon coupling strength $g(x)$:
\begin{align}
\!\!\!\dot{a}(t)\approx-i[\omega_c+\chi_{el}(t,\omega_c)]a(t) -\dfrac{\kappa}{2}a(t)-\sqrt{\kappa}b_{in}(t)\,,
\end{align}
where  $\kappa$ is the  decay rate of the cavity, $b_{in}(t)$ is the input signal that probes the cavity, while $\chi_{el}(t,\omega_c)=(1/2\pi)\int dt' \ e^{i\omega_ct'}\chi_{el}(t,t')$ with  \cite{KohlerPRL17}:
\begin{equation}
    \chi_{el}(t,t')=-i\theta(t-t')\langle[\mathcal{O}_{el}(t,t'),\mathcal{O}_{el}]\rangle\,,
    \label{susc}
\end{equation}
is the (time-dependent) electronic susceptibility associated with the operator $\mathcal{O}_{el}=\sum_{\sigma}\int dxg(x)\psi_{\sigma}^\dagger(x) \psi_{\sigma}(x)$ that couples to the photons. Moreover, $\mathcal{O}_{el}(t,t')\equiv U_{el}^\dagger(t,t')\mathcal{O}_{el}U_{el}(t,t')$, with $U_{el}(t,t')=\mathcal{T}\exp[-(i/\hbar)\int_{t'}^td\tau H_{el}(\tau)]$ being the evolution operator of the electronic system  ($\mathcal{T}$ is the time-ordering operator). Finally, $\langle\dots\rangle$ represents the expectation value with respect to the out-of-equilibrium electronic density matrix $\rho(t')$, whose derivation can, in general, be a formidable task. However, here we focus on adiabatic dynamics with respect to the topological gap $\Delta_t$, in which case we can assume the density matrix remains unaffected by the motion \cite{Albash_2012}.  

We start by analysing the case when $H_{el}$ is time-independent such that the topological nanowire harbors two distant MZMs, (e. g. due to an inhomogeneous chemical potential or magnetic field). They are described by left (right) fermionic operators $\gamma_{L(R)}$ that satisfy the Majorana condition $\gamma_{L(R)}^\dagger=\gamma_{L(R)}$ and encode a full fermionic state $c_M=(\gamma_L+i\gamma_R)/\sqrt{2}$ [$c_M^\dagger=(\gamma_L-i\gamma_R)/\sqrt{2}$] that is pinned at zero-energy in the absence of any interaction between the two end-modes. Its occupation, $n_M=c_M^\dagger c_M$,  determines the parity of the ground state defined as 
\begin{align}
\hat{\mathcal{P}}=1-2c_M^\dagger c_M=-2i\gamma_L\gamma_R\,,
\end{align}
whose expectation value, $\mathcal{P}\equiv\langle\hat{\mathcal{P}}\rangle$, can be used to describe the fidelity of the topological memory encoded in this two-dimensional subspace \cite{kitaev2001unpaired}. That is, $\mathcal{P}=\pm1$ in a state of given parity, corresponding to a state with even ($+$) or odd ($-$) number of electrons in the system.   
 
The electronic susceptibility can be written as $ \chi_{el}(\omega) =\chi_{el}^{MM}(\omega)+\chi_{el}^{BB}(\omega)+\chi_{el}^{BM}(\omega)$ \cite{dmytruk2015cavity}, being the sum of the contribution involving the MZMs only, the bulk states only, and the cross MZM-bulk states, respectively. Since photons mediate only parity conserving processes, $\chi_{el}^{MM}(\omega)=0$ and, moreover, for $\omega<2\Delta_t$, we can neglect also $\chi_{el}^{BB}(\omega)$. That is, in this frequency range, the susceptibility is dominated by the cross term (dropping the $BM$ superscript from here on), which reads:
\begin{align}
    \chi_{el}^\mathcal{P}(\omega)&\approx\sum_{n\in bulk}\frac{|\mathcal{M}_{nL}|^2+|\mathcal{M}_{nR}|^2+2\mathcal{P}{\rm Im}[\mathcal{M}_{nL}^*\mathcal{M}_{nR}]}{\omega-\epsilon_n+i\eta}\,,
    \label{susc}
\end{align}
where 
\begin{align}
\mathcal{M}_{n\alpha}&=\sum_{\sigma}\int dx \ g(x)\left[u^*_{\alpha\sigma}(x)u_{n\sigma}(x)-v^*_{\alpha\sigma}(x)v_{n\sigma}(x)\right]\nonumber\,,
\end{align}
are the matrix elements stemming from the local MZMs to the bulk states, with $u^*_{\alpha\sigma}(x)=v_{\alpha\sigma}(x)$ being the $\alpha=L,R$ Majorana electron and hole wave-functions, while $u_{n\sigma}(x)$ [$v_{n\sigma}(x)$] are the electron (hole) weights of the bulk state with energy $\epsilon_n$. Furthermore, $\eta$ is the quasiparticle linewidth assumed, for simplicity, to be the same for all states. The above expression holds for temperatures $T\ll\Delta_t$, so that the occupation of the bulk states $p_n\approx0$, and in the limit of negligible overlap between the MZMs, i.e. $\epsilon_M\approx0$ [for the full expression see Appendix~\ref{AppendixA}].  Eq.~\eqref{susc}, and in particular its imaginary part, ${\rm Im}\chi_{el}^\mathcal{P}(\omega)$ (quantifying the absorption), represents one of our central results. It shows that the electronic susceptibility that affects the photons is sensitive to the parity encoded by two non-overlapping MZMs via the {\it intensity} of the matrix elements with the extended bulk modes. For that to occur, the cavity field must cover both MZMs; otherwise, the two parities exhibit the same signal intensity. 

Since these effects are rooted in the interference of the MZMs with the extended bulk states, $\delta\chi_{el}(\omega)=|\chi_{el}^+(\omega)-\chi_{el}^-(\omega)|\propto1/L$ in the ballistic regime.  This is in contrast to the  detection approach scrutinised in Ref.~\cite{dmytruk2015cavity} which relied on the overlap of the MZMs wave-functions and resulted in a scaling $\delta\chi_{el}(\omega)\sim e^{-L/\xi}$, with $\xi=v_F/\Delta_t$ being the coherence length of the topological superconducting nanowire and $v_F$ is the Fermi velocity. We note that the power-law scaling should persist as long as the average energy level spacing of the bulk levels ($\delta\epsilon$), satisfies $\delta\epsilon\approx v_F/L>\eta$, which can be associated with a wire length $L^*=v_F/\eta$. Beyond this length, $\delta\chi_{el}(\omega)\sim e^{-L/L^*}$, which can be interpreted as follows: a bulk state injected at the left side has its amplitude reduced once reaching the other end because of its finite lifetime, therefore diminishing its common overlap with the two MZMs. Since typically $L^*\gg\xi$, there is a wide range of wire lengths for which these interference effects manifest such that $\epsilon_M\approx0$.   

To unravel the dynamical effects, let us assume the two MZMs are gliding rigidly, preserving the distance  $L$ between them. That can occur, for example, when the chemical potential that imprints the topological landscape in the wire obeys $\mu(x,t)=\mu(x-vt)$, with $v$ being the velocity of the rigid motion. Then, it is instructive to switch to the moving frame via a unitary transformation $\mathcal{W}(t)=\exp{(ivpt)}$ which results in the Hamiltonian $\widetilde{H}_{el}(v)=H_{el}(0)-vp$. When the coupling to the cavity is constant over the length of the nanowire, $g(x)\equiv g$, the transformation $\mathcal{W}(t)$ does not affect $H_{e-p}$, and the susceptibility acquires the same form as in Eq.~\eqref{susc}, but with the spectrum $\epsilon_n(v)$ and the matrix elements $\mathcal{M}_{n\alpha}$ now $v$-dependent according to $\widetilde{H}_{el}(v)$. To test these claims quantitatively, in the following we focus on a spin-orbit semiconducting nanowire (such as InAs and InSb) in proximity of an $s$-wave superconductor (such as Al) and subject to an external magnetic field. Such systems been recently under intense scrutiny in the quest of MZMs in engineered in semiconducting nanostructures \cite{mourik2012signatures,deng2012anomalous,das2012zero,churchill2013superconductor,deng2016majorana,de2018electric}.

\begin{figure}[t] 
\centering
\includegraphics[width=0.99\linewidth]{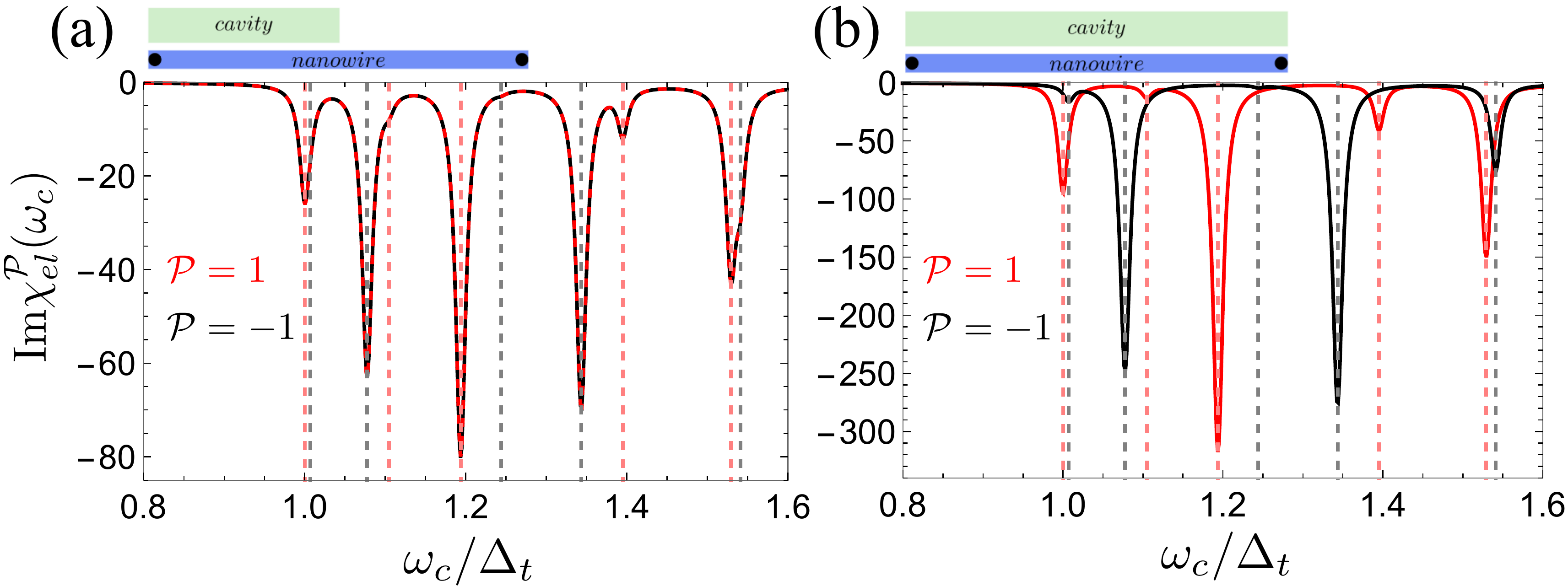}
\caption{Imaginary part of the susceptibility $\chi_{el}^\mathcal{P}(\omega)$ for parity $\mathcal{P}=\pm1$ as a function of the cavity frequency $\omega_c$ scaled with the topological gap $\Delta_t$.
(a) The cavity couples to $N_c = 50$ sites of the nanowire and (b) the cavity couples to the nanowire over its entire length $N_c \equiv N = 100$. The red (black) line corresponds to the even (odd) parity of the MZM. The dashed vertical lines indicate the first ten resonances. In (a) the intensity of the transitions is the same for both parities while in (b) the intensities differ substantially for $\epsilon_M\approx0$. The other parameters are fixed as $N=100$, $\Delta/t_h = 0.15$, $V_Z=2.5\Delta$, $\alpha/t_h = 0.4$, $\mu = 0$, and $\eta/t_h = 10^{-3}$.}
\label{fig:InhomogCoupling}
\end{figure}

\section{Lattice model}

The minimal tight-binding Hamiltonian describing a one-dimensional nanowire composed of $N$ lattice sites reads~\cite{rainis2013towards}
\begin{align}
H_{el}&(t) = 
 \sum_{\sigma,\sigma'}\sum_{j=1}^{N} c^\dag_{j,\sigma}\Big[\left(2t_h - \mu_j(t)\right) \delta_{\sigma\sigma'}
+ V_Z \sigma^x_{\sigma\sigma'}\Big]c_{j,\sigma'}\nonumber\\
&+\sum_{\sigma,\sigma'}\sum_{ j=1}^{N-1} 
c^\dag_{j+1,\sigma}T_{\sigma\sigma'}c_{j,\sigma'}+ \sum_{j=1}^{N} \Delta c^\dag_{j,\uparrow}c^\dag_{j,\downarrow}
+{\rm h. c.},
\label{eq:TightBindingH}
\end{align} 
where $c_{j\sigma}^\dag (c_{j\sigma})$ is the creation (annihilation) operator acting on electrons with spin $\sigma$ located at  site $j$, $\mu_j(t)$ is the position and time-dependent chemical potential,
$t_h = \hbar^2/\left(2 m a^2\right)$ is the hopping amplitude, with $m$ 
and $a$ being the effective mass and lattice constant, respectively. Here, $V_Z$ is the Zeeman energy, $T_{\sigma\sigma'}=\left(i \alpha \sigma^y_{\sigma\sigma'}-t_h \delta_{\sigma\sigma'} \right)$ is the total hoping matrix element, with $\alpha$ being the spin-orbit interaction (SOI) constant and $\sigma_{x,y,z}$ the Pauli matrices acting in the spin space, while $\Delta$ is the proximity-induced superconducting pairing potential. 

\begin{figure}[t] 
\centering
\includegraphics[width=0.99\linewidth]{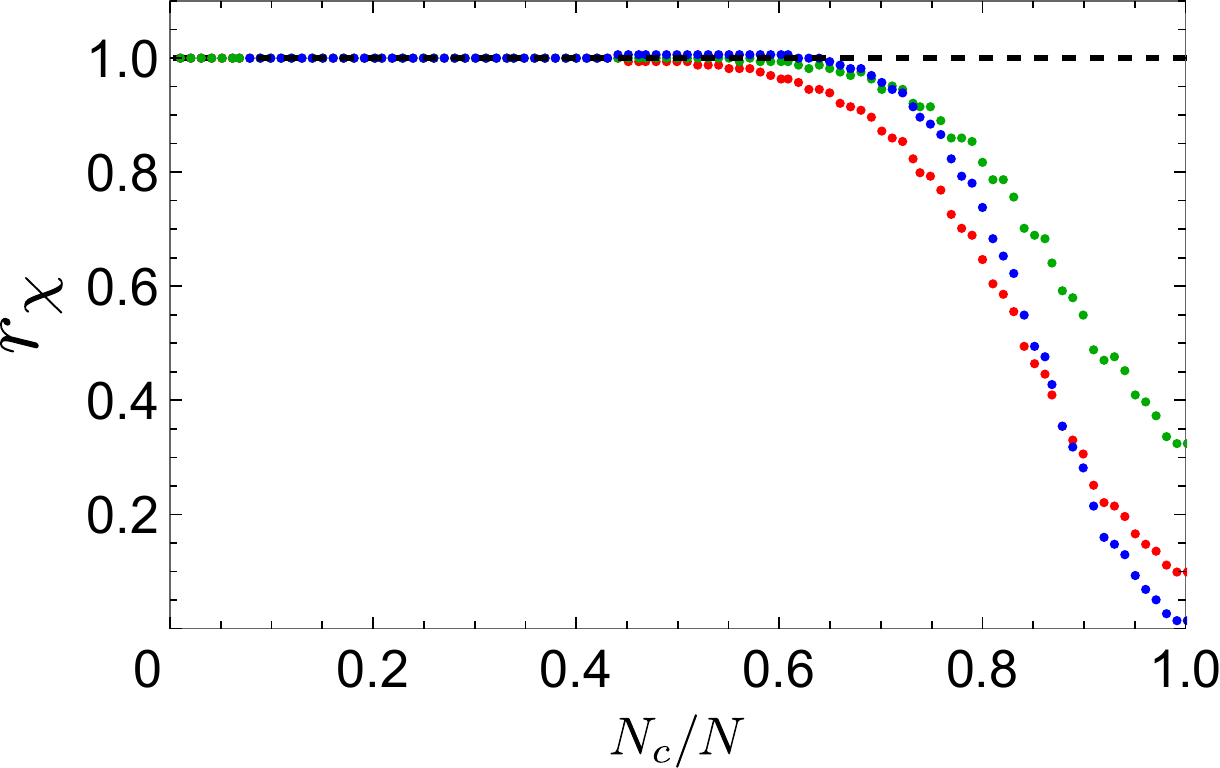}
\caption{Ratio of the $\mathcal{P}=\pm1$ susceptibilities, $r_\chi={\rm Im} \chi_{el}^{-}/{\rm Im}\chi_{el}^{+}$ ($ r_\chi={\rm Im} \chi_{el}^{+}/{\rm Im}\chi_{el}^{-}$) for ${\rm Im}\chi_{el}^-<{\rm Im}\chi_{el}^{+}$  (${\rm Im}\chi_{el}^+<{\rm Im}\chi_{el}^{-}$), as a function of the  fraction of sites coupled to the cavity, $N_c/N$. The red, green, and blue dots correspond to the first three absorption peaks.  All  other parameters are the same as in Fig.~\ref{fig:InhomogCoupling}.
}
\label{fig:Fig3Amplitude}
\end{figure}

We first consider a finite size nanowire subject to a time-independent and homogeneous chemical potential $\mu$, such that the topological gap $\Delta_t\equiv B-\sqrt{\Delta^2+\mu^2}>0$ and MZMs emerge at the ends of the nanowire. In our numerical simulations we focus on the strong SOI regime, $E_{SO} = m\alpha^2/\left(2\hbar^2\right)\gg V_Z,\Delta,\mu$, and a nanowire with $N$ lattice sites that result in vanishingly small zero mode splitting  ($\epsilon_M/\Delta_t = 2.4 \times 10^{-6}$). 
\begin{figure*}[t]
\centering
\includegraphics[width=0.99\linewidth]{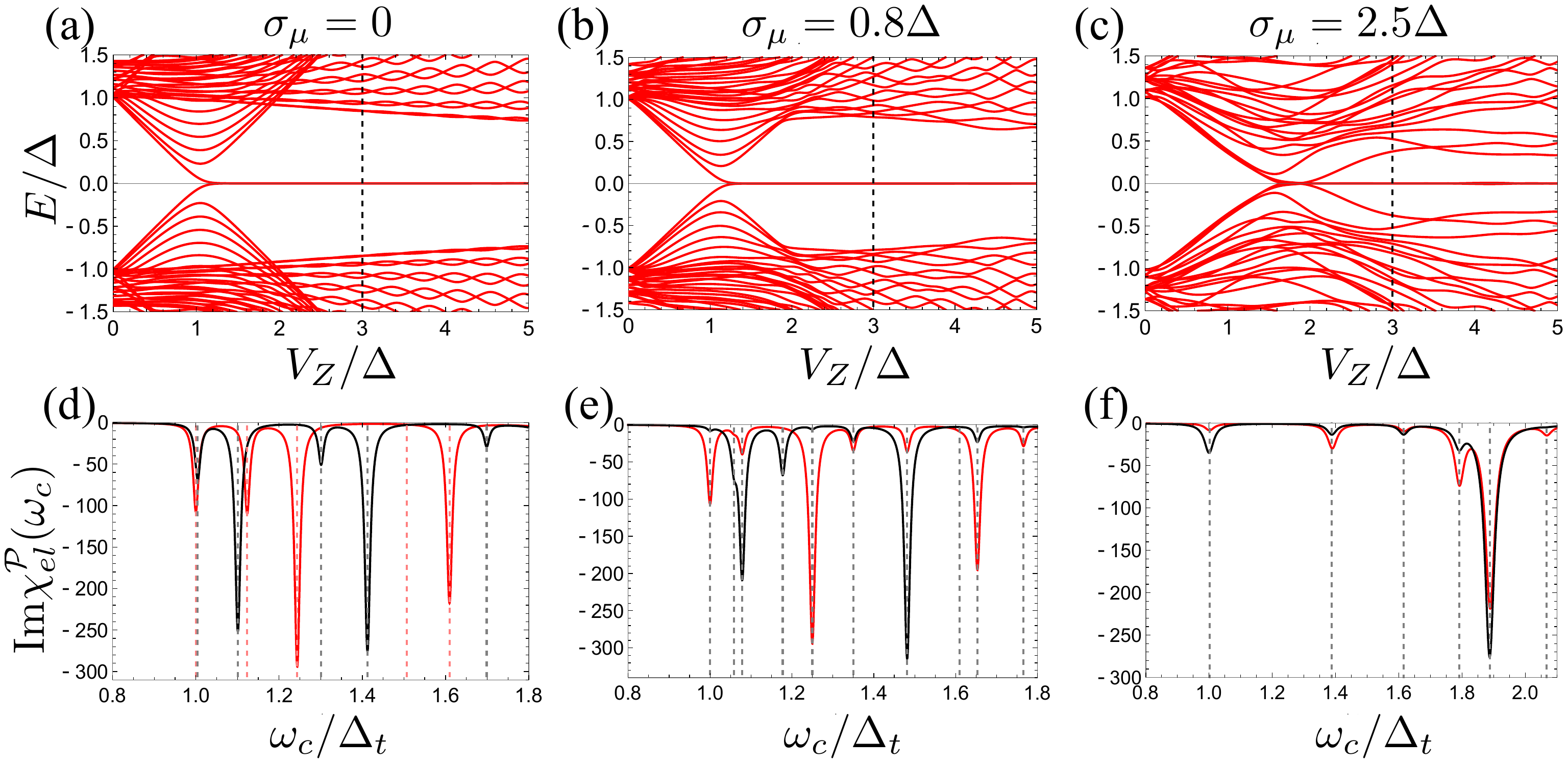}
\caption{(a) - (c) Energy spectrum of the finite-length disordered nanowire Eq.~\eqref{eq:TightBindingH} as a function of the Zeeman energy $V_Z/\Delta$ (the value of the standard deviation $\sigma_\mu$ is indicated directly in the plot). (d) - (f) Imaginary part of the electronic susceptibility $\text{Im}\chi_{el}^{\mathcal{P}}(\omega_c)$ as a function of the cavity frequency $\omega_c/\Delta_t$,
with the topological gap (d) $\Delta_t/\Delta = 0.88$, (e) $\Delta_t/\Delta = 0.79$, and (f) $\Delta_t/\Delta = 0.36$, for a fixed value of the Zeeman energy $V_Z/\Delta = 3$ 
(indicated by the black dashed line in panes (a) - (c)) and  disorder in the chemical potential with standard deviation $\sigma_\mu$. The other parameters are the same as in Fig.~\ref{fig:InhomogCoupling}. All cases show different absorption peaks for the two parities. }
\label{fig:EnergySusceptibilityDisorder}
 \end{figure*}
We evaluated the electronic susceptibility for different ratios $N_c/N$, with $N_c$ being the number of nanowire sites coupled to the microwave photons. When the photons interact with a fraction of the nanowire $N_c/N\leq1/2$, we found that there is no discernible difference in the cavity response between the two parities $\mathcal{P}=\pm1$, or  $\delta\chi_{el}(\omega)=0$ [see Fig.~\ref{fig:InhomogCoupling}~(a)]. However, by increasing $N_c$ towards $N$,  the two parities start exhibiting different intensity patterns, with the largest deviations between the two occurring  when the cavity is coupled to the entire nanowire. This effect, which stems from the last terms in Eq.~\eqref{susc}, is depicted in  Fig.~\ref{fig:InhomogCoupling}~(b), where we plot the evolution of the imaginary part of the susceptibility, ${\rm Im}\chi_{el}^\mathcal{P}$, with the cavity frequency $\omega_c$. We note that the same features are exhibited by the real part of the susceptibility which, however, are relegated to Appendix~\ref{AppendixB}. To further illustrate the difference in the susceptibility for the two parities, in Fig.~\ref{fig:Fig3Amplitude} we show the ratio between the even- and odd-parity susceptibility, $r_\chi={\rm Im}\chi_{el}^-/{\rm Im}\chi_{el}^+$, as a function of $N_c/N$. Since the $r_\chi<1$ if more than half of the nanowire is coupled to the cavity, tuning the inhomogeneous coupling between the nanowire and the cavity allows to probe the non-locality of the MZMs. In fact, such discrimination should persist even for more pairs of non-overlapping MZMs living in the same topological material. Additionally, such a mechanism could be utilised to initialise the nanowire in a given parity state without the requirement to fuse the two MZMs by bringing them in close proximity to each other.

In order to test the robustness of the susceptibility, we have analyzed whether disorder in the chemical potential hampers the discrimination between the parities. Here, we do not perform ensemble averaging, but instead use  quenched spatial disorder originating from fixed random configuration chosen from a Gaussian distribution, consistent with the experiment being performed at very low temperatures \cite{PanPRB21}.  In Fig.~\ref{fig:EnergySusceptibilityDisorder} (a) - (c) we show the energy spectrum of Eq.~\eqref{eq:TightBindingH} as a function of the Zeeman energy for three different strengths of on-site disorder in the chemical potential, with the standard deviation $\sigma_\mu$. We note that zero-energy MZMs are present even for $\sigma_\mu > \Delta_t$, but with the modified bulk energy levels [see Fig.~\ref{fig:EnergySusceptibilityDisorder}~(c) ]. Fixing the value of the Zeeman energy $V_Z/\Delta = 3$ we plot the imaginary part of the susceptibility ${\rm Im}\chi_{el}^{\mathcal{P}}$ as a function of the cavity frequency $\omega_c$ for different disorder strengths in Fig.~\ref{fig:EnergySusceptibilityDisorder}~(d)-(f). While all the matrix elements that enter in the susceptibility are affected, we see that the two parities still exhibit significantly different intensity pattern when the on-site fluctuations in $\mu$ are smaller than the topological gap $\Delta_t$ [see Fig.~\ref{fig:EnergySusceptibilityDisorder}~(d)-(e)], while the difference in amplitudes still persists even for large values of disorder with $\sigma_\mu>\Delta_t$ [see Fig.~\ref{fig:EnergySusceptibilityDisorder}~(f)]. Therefore, the susceptibility remains a good quantity to harness for distinguishing between the even and the odd parities even in the presence of disorder. Our results are also consistent with previous works that demonstrate that for moderate disorder (strength smaller than $\Delta_t$) the MZMs are largely unaffected \cite{BrouwerPRL11,BrouwerPRB11,LobosPRL12,SauPRB13}. 

\begin{figure}[t] 
\includegraphics[width=0.99\linewidth]{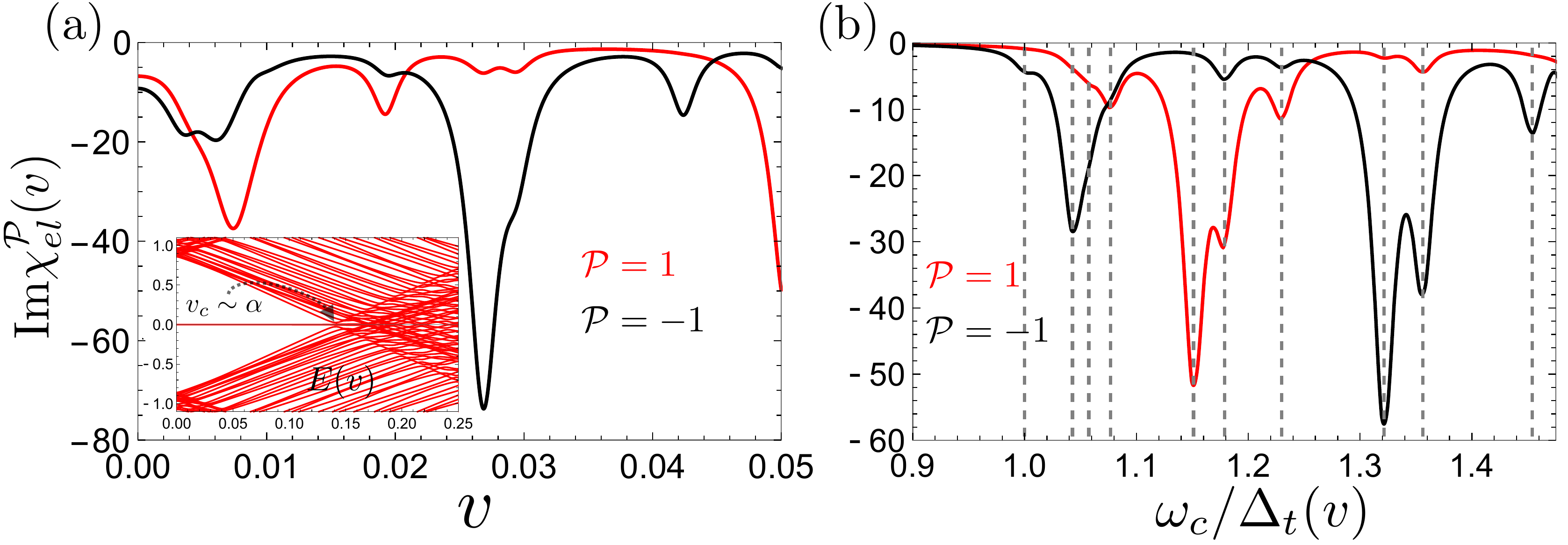} 
\caption{(a) Imaginary part of the susceptibility ${\rm Im}\chi_{el}^\mathcal{P}$ of the ring as a function of velocity $v$ for $\omega_c/\Delta = 0.9$. The red (black) line describes the even (odd) Majorana parity. 
Inset: Energy spectrum of the ring in the moving frame as a function of $v$, becoming gapless at $v\sim\alpha$. (b) Imaginary part of the susceptibility ${\rm Im}\chi_{el}^\mathcal{P}$ of the ring as a function of the cavity frequency normalized by the effective gap $\omega_c/\Delta_{t}(v)$ for $v =0.05$ [$\Delta_{t}(v)=0.59\Delta$]. The red (black) line describes the even (odd) parity. The gray dashed lines indicate the position of absorption peaks.
For all curves we used $V_Z/\Delta = 2.5$, $N_{top}=100$, $N_{total}=300$, $\Delta/t_h = 0.15$, $\alpha/t_h = 0.4$, $\mu/t_h = 1.5$, and $\eta/t_h = 10^{-3}$.
} 
\label{fig:DynamicsRotated}
\end{figure}

To investigate the MZM dynamics, we consider a nanowire in a ring geometry that is subject to a chemical potential $\mu(j, t)$ that glides with velocity $v$, such that a fraction of the ring is in the topological phase, supporting MZMs ($\Delta_t>0$), while the rest of the ring remains in the trivial phase ($\Delta_t<0$). To account for the gliding in the moving frame, we substitute $t_h\rightarrow t_he^{imav}$ in the tunneling term in Eq.~\eqref{eq:TightBindingH}. The resulting energy spectrum of the ring in the moving frame as a function of $v$ is depicted in the inset of Fig.~\ref{fig:DynamicsRotated}~(a). It shows that the effective gap in the energy spectrum, $\Delta_{t}(v)$, decreases with increasing the velocity $v$, with the system becoming gapless at a critical $v_c\approx\alpha$ \cite{ShnirmanPRB13}. Nevertheless, the MZMs remain localised near zero energy up to $v_c$.
Next, we scrutinise the evolution of the absorption peaks with the MZM dynamics. In Fig.~\ref{fig:DynamicsRotated}~(a) we show the imaginary part of the susceptibility as a function of velocity $v$, exhibiting  an oscillatory peak structure, similar to the dependence on $\omega_c$ in the static case. Here, the gliding of the chemical potential affects the bulk states wave-function oscillation pattern (for a given energy $\omega_c$), modifying their overlap with the MZMs, whose wave-functions instead remain largely unaffected by $v$ in the adiabatic limit. 
In Fig.~\ref{fig:DynamicsRotated} (b) we show the frequency-dependence of ${\rm Im}\chi_{el}^\mathcal{P}(\omega_c)$ at finite $v$. Interestingly, as compared to the static case, each resonance peak is now split into two. This feature originates from an asymmetry between the left and right moving bulk states induced by the gliding motion of $\mu(j,t)$, absent  at $v=0$, and which alters their constructive or destructive interference.  

\section{Low-energy model}

We can capture the essentials of the interference pattern  with a low-energy continuous model describing the nanowire. The second-quantized expression for the nanowire Hamiltonian in the continuum limit is
\begin{align}
    \mathcal{H}_{w}(t)=\frac{1}{2}\int dx\Psi^\dagger(x)H_{BdG}(x,t)\Psi(x)\,,
\end{align}
that it is written in terms of the Nambu spinors $\Psi(x)=\{\psi_\uparrow(x),\psi_\downarrow(x), \psi^\dagger_\downarrow(x), -\psi^\dagger_\uparrow(x)\}^T$, and with the (time-dependent) BdG Hamiltonian \begin{align}
    H_{BdG}=\left(\frac{p^2}{2m}-\mu(x,t)+\alpha p\sigma_z\right)\tau_z+B\sigma_x+\Delta\tau_x\,,
    \label{low}
    \end{align}
where $\mu(x,t)\equiv\mu_0+\delta\mu(x,t)$ is the time and the position dependent chemical potential that eventually induces the motion of the MZMs. Let us define  $\Delta_{t}(x,t)=\sqrt{\Delta^2+\mu^2(x,t)}-B$ as the inhomogeneous topological gap, and $\mu(x,t)=\mu_0+\delta\mu(x,t)$ such that $\Delta_{t}=0$ for $\delta\mu(x,t)=0$ (determining the positions of the MZMs). While describing the finite superconducting nanowire analytically is in general a difficult task, in the limit of strong SOI, quantified by ${\rm max}[\Delta_{t}(x,t)]\ll m\alpha^2\Delta/\mu_0 B$, we can make progress following the expositions of Refs.~\cite{KarzigPRX13,ShnirmanPRB13}. First,  when the system is near the topological phase transition the minimum gap occurs at $p=0$ and we can neglect the quadratic term in Eq. \eqref{low}. Second,  assuming the topological region moves rigidly at finite velocity $v$, the pairing parameter becomes $\delta\mu(x)\rightarrow \delta\mu(x-vt)$, and  we can use a unitary transformation $U(t)=e^{-ipvt}$ to gauge away the time-dependence. This results in the following  Hamiltonian in the moving frame:
\begin{align}
    \widetilde{H}^l_{BdG}(x)=u p\Sigma_z+\Delta_{t}(x,0)\Sigma_x-vp\,,
\end{align}
with $u=\alpha\sqrt{1-(\mu_0/B)^2}$, $p=-i\hbar\partial/\partial x$,  and $\Delta_{t}(x)\approx\mu_0\delta\mu(x)/B$ (the corresponding high-energy Hamiltonian describes bulk states with energies larger than $\epsilon(p)>2B$ and are disregarded from here on \cite{KarzigPRX13}). The Pauli matrices ${\bs \Sigma}=(\Sigma_x, \Sigma_y, \Sigma_z)$ act in the space spanned by $\{(-a_+,-a_-,-a_-,a_+), (-a_-,-a_+,a_+,-a_-)\}$, with $a_{\pm}=\sqrt{1\pm\Delta/B}$ \cite{KarzigPRX13}.  When the system is homogeneous, i. e. $\Delta_{t}={\rm const}$, the eigen-energies become:
\begin{align}
    \epsilon_{\pm}(p)=-vp\pm\sqrt{(pu)^2+\Delta_{t}^2}\,,
\end{align}
which corresponds to a tilted spectrum with a topological gap $\Delta_t(v)=\Delta_t(0)\sqrt{1-\beta^2}$, where $\beta=v/u$. The system becomes gapless at $v=u$, consistent with our findings in the previous section [see inset of Fig.~\ref{fig:DynamicsRotated}(a)]. The capacitive coupling of the electrons to the cavity effectively amounts to changes in the chemical potential $\mu_0$, and thus acts in the low-energy sector as $H_{e-p}^l\approx g(\mu_0/B)(a^\dagger+a)\Sigma_x$. In order to obtain the electronic susceptibility in Eq.~\eqref{susc}, we need to determine the wave-functions of both the MZMs and the bulk states, respectively. For that, we utilise a scattering approach assuming the topological gap profile $\Delta_{t}(x)=\Delta_{t}$ for $x\in(-\infty,0){\rm U}(L,\infty)$, and $\Delta_{t}(x)=-\Delta_{t}$ otherwise. Consequently, in each region we can utilise linear combinations of the homogeneous space solutions, and then invoke continuity of the wave-functions at $x=0$ and $x=L$. In a given region, the states with a given energy $\omega$ read:
\begin{align}
    \psi_{\pm}(\omega,x)=\frac{1}{\mathcal{N}_{\pm}}\left(1,C_\beta(\omega\mp\sqrt{\omega^2-1})\right)^Te^{ik_{\pm}x}\,,
\end{align}
where $C_\beta={\rm sign}(\Delta_{t})\sqrt{(1-\beta)/(1+\beta)}$ is a contraction factor, $k_{\pm}=(\beta \omega\pm\sqrt{\omega^2-1})/(u\sqrt{1-\beta^2})$ are the two momenta for a given energy $\omega$ (for in-gap states being complex), while $\mathcal{N}_{\pm}=\sqrt{1+C_\beta^2(\omega\mp\sqrt{\omega^2-1})^2}$ are normalization factors. All the energies above are expressed in terms of the effective gap $\Delta_{t}(v)$. While the scattering problem can be solved exactly for arbitrary $v$, the resulting expressions are lengthy and uninspiring. Instead, by solving the boundary conditions detailed in Appendix~\ref{AppendixC}, we can extract simple analytical expressions valid in the limit $\beta\ll1$ and $\epsilon_M=0$ ($L\gg u/\Delta_t$): 
\begin{align}
    {\rm Im}\chi^\mathcal{P}_{el}&\approx\left\{
    \begin{array}{cc}
    a(k)\cos{\frac{(k-k_v)L}{2}}+b(k)\sin\frac{(k-k_v)L}{2}\,, & \mathcal{P}=1\\\\
    a(k)\sin{\frac{(k-k_v)L}{2}}-b(k)\cos\frac{(k-k_v)L}{2}\,, & \mathcal{P}=-1
\end{array}
\right.\label{app_chi}\nonumber\\\\
    &a(k)=\frac{16k^3\sin(kL)}{(1+k^2)^2[k^4+4(1-k^2)\sin^2(kL)]}\nonumber\,,\\
    &b(k)=\frac{16k^2[\sin(kL)-2k\cos{(kL)}]}{(1+k^2)^2[k^4+4(1-k^2)\sin^2(kL)]}\,,
\end{align}
where $k=\sqrt{\omega^2-1}/u$, while $k_v=\beta\sqrt{\omega^2-1}/u$ is the momentum boost stemming from the gliding. \begin{figure}[t] 
\includegraphics[width=0.99\linewidth]{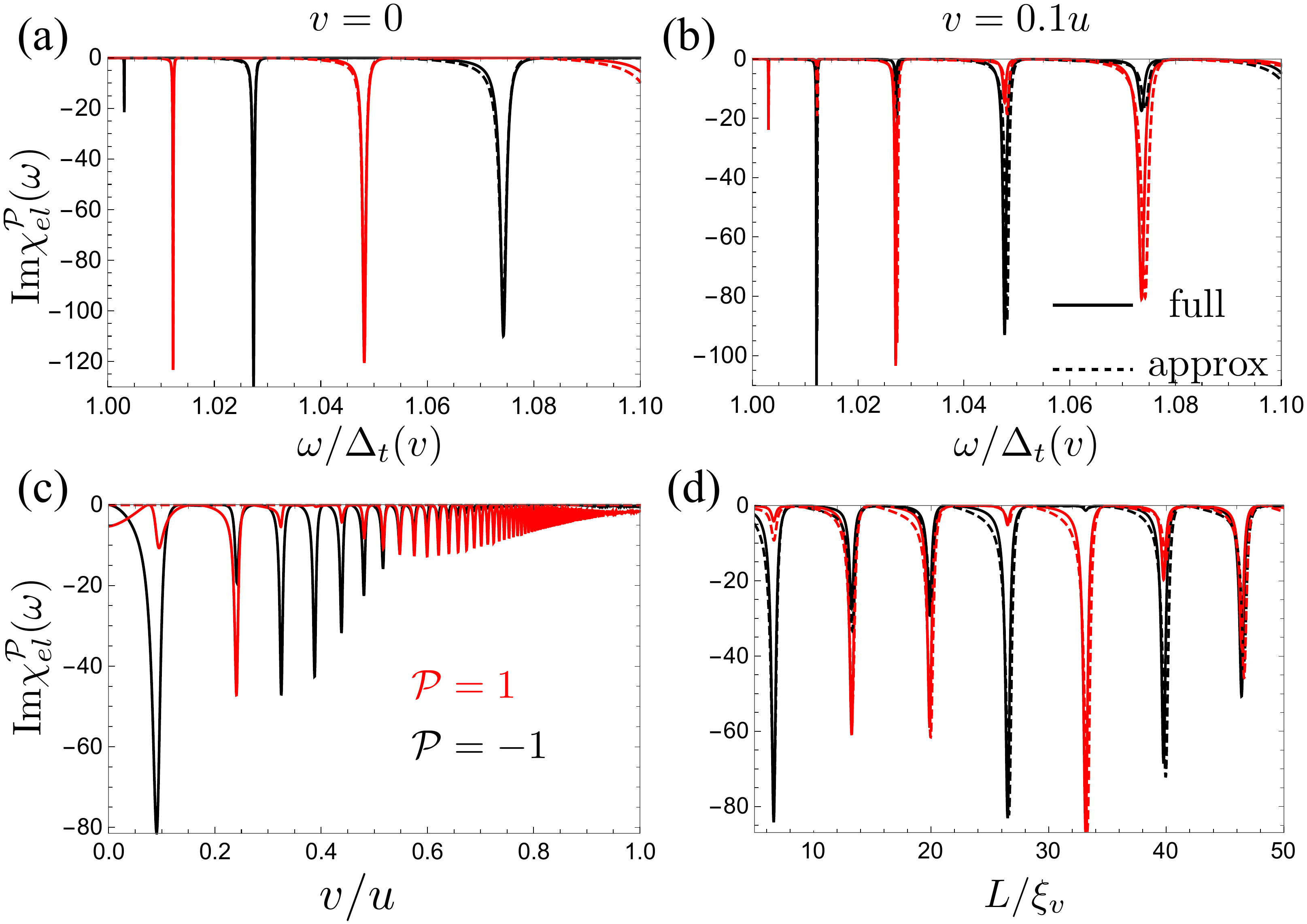}
\caption{The imaginary part of the susceptibility evaluated from the effective low-energy model. (a)   Frequency dependence of the susceptibility for the two parities at $v=0$. (b) same at $v=0.1$  (c) $v$-dependence for $\omega=1.1\Delta_t(0)$. (d) Dependence on the size of the topological region $L$ scaled with the coherence length $\xi_v=1/\Delta_t$ at $v=0.1$. The full lines correspond to the exact result from the analytic model, while the dashed lines pertain to the approximate expressions in Eq.~\eqref{app_chi}. In (a), (b), and (c) we used $L=40$, with $u=1$ in all plots.}
\label{fig:Fig5}
\end{figure}
We see that for $v=0$, ${\rm Im}\chi_{el}^{+(-)}(\omega)\propto \sin^4(kL/2)[\cos^4(kL/2)]$,  which shows that when one parity intensity is maximum, the other parity exhibits a vanishing absorption signal, in qualitative agreement with the findings from the lattice model. In Fig.~\ref{fig:Fig5} we show the results for the susceptibility obtained by solving the full scattering model (from the boundary conditions), against the approximate results in Eq.~\ref{app_chi} extracted for energies close to the band edge $\omega\geq\Delta_t(v)$. We see very good agreement between the two in the adiabatic limit $\beta\ll1$. A finite $v$ breaks this perfect alternation of maxima by inducing a  momentum boost $k_v=\beta\sqrt{\omega^2-1}/u$ ($k_v=-\beta\sqrt{\omega^2-1}/u$) to the left (right) moving bulk states, affecting the interference pattern, consistent again with the numerical findings. Put differently, the extended bulk states facilitate a momentum-resolved microwave spectroscopy of the localised MZMs with the gliding acting as to modify the resulting interference pattern via a momentum boost $k_v$ \cite{AuslaenderScience02,TserkovnyakPRL02}.

\section{Conclusions}

To summarize, in this work we have shown that photons in a microwave cavity coupled to a topological superconducting nanowire that harbors MZMs are affected by both the parity of the ground state and the MZMs dynamics. These effects originate from the interference between the localised MZMs and the extended bulk states in the presence of long-range photons, and do not require any overlap of the MZMs wave-functions. Auspicious developments in using MZMs in nanowires for quantum computing are currently hampered by the difficulty in distinguishing them from other trivial bound states. Utilising long-range photons to map out the non-local structure of MZMs and their dynamics in semiconducting  nanowires could offer complementary avenues to the local tunneling spectroscopy to elucidate this conundrum.

\section*{Acknowledgments} 

This project was supported by the Foundation for Polish Science through the International Research Agendas program co-financed by the European Union within the Smart Growth Operational Programme (MAB/2017/1), and by the National Science Centre (Poland) OPUS 2021/41/B/ST3/04475 (MT). This project has received funding from the European Union’s Horizon 2020 research and innovation programme under the Marie Skłodowska-Curie Grant Agreement No. 892800 (OD).

\appendix

\section{Input-output framework}\label{AppendixA}

The single-mode cavity couples capacitively to the electronic density:
\begin{align}
H_{e-p}&=\mathcal{O}_{el}(a^\dagger+a)\,,
\end{align}
where 
\begin{align}
\mathcal{O}_{el}=\sum_{\sigma}\int dx \ g(x)\psi_{\sigma}^\dagger(x) \psi_{\sigma}(x)\,,
\end{align}
with $g(x)$ being the coupling strength to the electron density $\psi^\dagger_{\sigma}(x)\psi_{\sigma}(x)$ at position $x$, and $a$ ($a^\dagger$) is the photon annihilation (creation) operator. The equation of motion for the photonic operator reads~\cite{clerk2010introduction}
\begin{align}
\dot{a}(t)&=-i\omega_c a(t)-i\mathcal{O}_{el,H}(t)-\dfrac{\kappa}{2}a(t)-\sqrt{\kappa}b_{in}(t)\,,
\end{align}
where $\kappa$ is the decay rate of the cavity field into the external transmission line, $b_{in}(t)$ is the input field impinging onto the cavity, while $\mathcal{O}_{el,H}(t)$ is evolved by the total Hamiltonian (Heisenberg picture). The input and output fields (which is eventually measured in an experiment) are related via:
\begin{align}
b_{out}(t)=b_{int}(t)+\sqrt{\kappa}a(t)\,.
\end{align}
Following Ref.~\cite{KohlerPRL17}, we can write for the total density matrix  $\dot{\rho}(t)=\mathcal{L}(t)\rho(t)-i[\mathcal{O}_{el},\rho(t)](a^\dagger+a)$, with $\mathcal{L}(t)$ being the Liouvillean of the electronic system only (but including intrinsic dissipation channels), while in the absence of the cavity the density matrix obeys $\dot{\rho}_{0}(t)=\mathcal{L}(t)\rho_0(t)$. Then, in leading order in the coupling to the photons, we have:
\begin{align}
    &\langle\mathcal{O}_{el}(t)\rangle=\langle\mathcal{O}_{el}(t)\rangle_0+\int dt'\chi_{el}(t,t')[a(t')+a^\dagger(t')]\,,\label{average}\nonumber\\
    &\chi_{el}(t,t')=-i{\rm Tr}[[\mathcal{O}_{el}(t,t'), \mathcal{O}_{el}]\rho_0(t')]\theta(t-t')\,,
\end{align}
with the latter being the time-dependent electronic susceptibility measured by the cavity. Also,  $\mathcal{O}_{el}(t,t')\equiv U_{el}^{\dagger}(t,t')\mathcal{O}_{el}U_{el}(t,t')$, with $U_{el}(t,t')$ being the propagator describing the unitary evolution of the system,  while $\langle\dots\rangle_0$ means average with respect to $\rho_0(t)$. In Eq.~\eqref{average}  we can write $a(t')\approx a(t)e^{i\omega_c(t-t')}$, which disregards higher  order corrections in the coupling to the electrons. Finally, all this allow us to write the following local in time equation for the evolution of the photonic field:
\begin{align}
\dot{a}(t)&\approx-i\omega_ca(t)-i\langle \mathcal{O}_{el}(t)\rangle_0-ia(t)\chi_{el}(t,\omega_c) \notag\\
&-\dfrac{\kappa}{2}a(t)-\sqrt{\kappa}b_{in}(t)\,,
\end{align}
where we introduced the (time-dependent) susceptibility
\begin{align}
    \chi_{el}(t,\omega_c)=-i\int_{-\infty}^{\infty}dt' e^{i\omega_c t'}\chi_{el}(t,t')\,.
\end{align}
When the input field $b_{in}\gg\langle \mathcal{O}_{el}(t)\rangle_0$, we can neglect the latter from the equation of motion, resulting the expression showed in the manuscript. 

In the main text, we are concerned with the dynamics induced by a gliding chemical potential $\mu(x-vt)$. A unitary transformation $\mathcal{W}(t)=\exp{(iv\,p\,t)}$ renders the Hamiltonian time-independent
\begin{align}
    &\widetilde{H}_{el}(t)=
    \mathcal{W}^\dagger(t)H_{el}(t)\mathcal{W}(t)-i\mathcal{W}^{\dagger}(t)\dot{\mathcal{W}}(t)\notag\\
    &\equiv H_{el}(0)-vp\,,
\end{align}
while the eigenstates in the lab frame and the moving frame are related by $|\psi_n(t)\rangle=\mathcal{W}(t)|\widetilde{\psi}_n(t)\rangle$. The electronic evolution operator is
\begin{align}
    U_{el}(t,t')=\mathcal{W}^{\dagger}(t)e^{-i\widetilde{H}_{el}(t-t')}\mathcal{W}(t')\,,
\end{align}
and the susceptibility becomes:
\begin{align}
\chi_{el}(t,t')&=-i{\rm Tr}[[\mathcal{O}_{el}(t,t'), \mathcal{O}_{el}(t')]\tilde{\rho}_0(t')]\theta(t-t')\nonumber\,,
\end{align}
where $\mathcal{O}_{el}(t,t')=e^{i\widetilde{H}_{el}(t-t')}\mathcal{O}_{el}(t)e^{-i\widetilde{H}_{el}(t-t')}$, with $\mathcal{O}_{el}(t)=\mathcal{W}^\dagger(t)\mathcal{O}_{el}\mathcal{W}(t)$ and $\tilde{\rho}_0(t)=\mathcal{W}(t)\rho_0(t)\mathcal{W}^\dagger(t)$. When the cavity field is constant  over the length of the nanowire (i.e. $g(x)\equiv g$), the displacement operator does not affect the coupling Hamiltonian $H_{e-p}$. Therefore, the combined system remains time-independent in the moving frame. Assuming adiabatic motion, we can safely assume the dynamics does not affect the density matrix, which is given by $\tilde{\rho}_0(t)\equiv\tilde{\rho}_0=\sum_{n}p_n|\widetilde{\psi}_n\rangle\langle\widetilde{\psi}_n|$, with $p_n$ being the occupation of the BdG levels, and which makes the susceptibility dependent only on the time difference $t-t'$. Consequently,  $\chi_{el}(t,\omega_c)\equiv\chi_{el}(\omega_c)$, allowing  to determine the susceptibility analogously to the time-independent case \cite{dmytruk2015cavity}.

\section{Derivation of the electronic susceptibility}

It is instructive to express the fermionic operators as follows
\begin{align}
    \psi^\dagger_\sigma(x)&=\sum_{n}\left(u^*_{n\sigma}(x)c_n^\dagger+v_{n\sigma}(x)c_n\right)\,,
\end{align}
where $u_{n\sigma}(x)$ ($v_{n\sigma}(x)$) are the electron (hole) coherence factors for state $n$ at position $x$ and with spin $\sigma$, while $c_{n}$ ($c_n^\dagger$) are the annihilation (creation) operator for the bogoliubons of energy $\epsilon_n$ that diagonalise the electronic Hamiltonian in the moving frame, i. e. $\widetilde{H}_{el}=\sum_{n}\epsilon_n(c_n^\dagger c_n-1/2)$.  Then, by employing the equation of motion $e^{i\widetilde{H}_{el}t}c_ne^{-i\widetilde{H}_{el}t}=c_ne^{-i\epsilon_nt}$ we can insert the above fields in the definition of $\mathcal{O}_{el}$ to find
\begin{align}
    &\chi_{el}(\omega)=\sum_{n,m}\sum_{\sigma,\tau}\int dx\ dx'[u^*_{n\sigma}(x)u_{m\sigma}(x)-v_{m\sigma}(x)v^*_{n\sigma}(x)]\notag\\
    &\times[u^*_{m\tau}(x')u_{n\tau}(x')-v_{n\tau}(x')v^*_{m\tau}(x')]\frac{p_n-p_m}{\omega+\epsilon_n-\epsilon_m+i\eta}\nonumber\\
    &+\frac{1}{2}[u^*_{n\sigma}(x)v^*_{m\sigma}(x)-u^*_{m\sigma}(x)v^*_{n\sigma}(x)]\nonumber\\
    &\times[v_{n\tau}(x')u_{m\tau}(x')-v_{m\tau}(x')u_{n\tau}(x')]\frac{1}{\omega+\epsilon_n+\epsilon_m+i\eta}\nonumber\\
    &-\frac{1}{2}[v_{n\sigma}(x)u_{m\sigma}(x)-v_{m\sigma}(x)u_{n\sigma}(x)]\nonumber\\
   &\times [u^*_{n\tau}(x')v^*_{m\tau}(x')-u^*_{m\tau}(x')v^*_{n\tau}(x')]\frac{1-p_m-p_n}{\omega-\epsilon_n-\epsilon_m+i\eta}\nonumber\,,
\end{align}
where $\eta$ is a small positive rate that encodes the lifetime of the levels. In this work, we are interested in the imaginary part of the susceptibility, and more specifically on energies $\omega<2\Delta_t$. That is, we only account for the contributions that involve the bulk states and the MZMs:
\begin{align}
    \chi_{el}(\omega)&\approx\sum_m\bigg(|\mathcal{M}_{mM}^o|^2\frac{p_M-p_m}{\omega+\epsilon_M-\epsilon_m+i\eta}\notag\\
    &+|\mathcal{M}_{mM}^e|\frac{1-p_m-p_M}{\omega-\epsilon_M-\epsilon_m+i\eta}\bigg)\,,
\label{susc_append}    
\end{align}
where
\begin{align}
    \mathcal{M}_{mM}^e&=\sum_{\sigma}\int dx\left[u^*_{M\sigma}(x)u_{m\sigma}(x)-v^*_{M\sigma}(x)v_{m\sigma}(x)\right]\,,\\
    \mathcal{M}_{mM}^o&=\sum_{\sigma}\int dx\left[v_{M\sigma}(x)u_{m\sigma}(x)-u_{M\sigma}(x)v_{m\sigma}(x)\right]\nonumber\,,
\end{align}
are the intensities of the matrix elements describing the transitions into the bulk for the odd and even parities, respectively. Expressing the electron and hole weights of the MZM as:
\begin{align}
    u_{M\sigma}(x)&=\frac{1}{\sqrt{2}}[u_{L}^\sigma(x)+iu_{R}^\sigma(x)]\nonumber\,,\\
    v^*_{M\sigma}(x)&=\frac{1}{\sqrt{2}}[u_{L}^\sigma(x)-iu_{R}^\sigma(x)]\,,
\end{align}
readily allows us to identify the left and the right Majorana wave-functions, that are separated in space and satisfy the Majorana condition $u_{L(R)\sigma}(x)=v^*_{L(R)\sigma}(x)$. With that, we can finally write:
\begin{align}
    |\mathcal{M}_{mM}^{e,o}|^2=|\mathcal{M}_{mL}|^2+|\mathcal{M}_{mR}|^2\pm2{\rm Im}[\mathcal{M}_{mL}^*\mathcal{M}_{mR}]\nonumber\,, 
\end{align}
where $\mathcal{M}_{m\alpha}=\sum_{\sigma}\int dx\left[u^*_{\alpha\sigma}(x)u_{m\sigma}(x)-v^*_{\alpha\sigma}(x)v_{m\sigma}(x)\right]$ are the matrix elements stemming from the local Majorana $\alpha=L,R$ states and the positive energy bulk modes. The imaginary part of the susceptibility, which is responsible for the absorption peaks,  reads:
\begin{align}
    {\rm Im}\chi_{el}^{e,o}(\omega)&=-\pi\sum_m\Big[|\mathcal{M}_{mL}|^2+|\mathcal{M}_{mR}|^2\notag\\
    &\pm2{\rm Im}[(\mathcal{M}_{mL})^*\mathcal{M}_{mR}]\Big]\delta(\omega-\epsilon_m)\,,
\end{align}
where we assumed the low temperature regime compared to the topological gap $\Delta_t$ ($p_m\approx0$) and negligible Majorana overlap ($\epsilon_M\approx0$). This shows that while the resonance positions are the same for the two parities their intensity differ, but only if the cavity couples to both MZMs.

\section{Numerical approach for the Rashba nanowire lattice model} \label{AppendixB}

In this section, we provide details on the numerical evaluation of the electronic susceptibility for the tight-binding model given by Eq.~\eqref{eq:TightBindingH} in the main text. To emulate the continuum analogue of the gliding chemical potential, $\mu(x,t)=\mu(x-vt)$, we perform the substitution
\begin{equation}
    t_h\rightarrow t_he^{imav/\hbar}\,,
\end{equation}
in the tight-binding model in Eq.~\eqref{eq:TightBindingH} in the moving frame.  We are interested in cases when $\hbar/(ma)\approx v_F\gg v$, so that we can write $t_he^{imav/\hbar}\approx t_h(1+imav/\hbar)$. Therefore, the static electronic Hamiltonian is supplemented with the term
\begin{align}
    H_v=-i\frac{\hbar v}{2a}\sum_{j,\sigma}(c^\dagger_{j+1\sigma}c_{j\sigma}-c^\dagger_{j\sigma}c_{j+1\sigma})\,.
\end{align}
Diagonalising the full Hamiltonian, including the term above, results in the spectrum presented in Fig.~\ref{fig:DynamicsRotated} in the main text that exhibits a gap closing at $v\approx\alpha$. For the one-dimensional ring with $N$ sites Eq.~\eqref{eq:TightBindingH} the electronic susceptibility expressed in the moving frame reads
\begin{align}
        \chi_{el}(t-t')&=-i\theta(t-t'){\rm Tr}[[n(t), n(t')]\tilde{\rho}_0(t')]\,,
        \label{eq:SusceptibilityTightBinding}
\end{align}
where $n=\sum_{j=1}^{N}c_j^\dag c_j$ and $n(t)$ is the electron density operator in the interaction picture evolved with the total Hamiltonian, including $H_v$ above. The density matrix was chosen as $\tilde{\rho}_0=p_M|0\rangle\langle0|+(1-p_M)|1\rangle\langle1|$, with $p_M=0,1$ and $|0,1\rangle$ being the occupation and the many-body state of parity $\mathcal{P}=-1,1$, respectively (we further assumed zero temperature, $T=0$, so that all negative bulk levels are occupied). 

We can make connection to the continuum model and write $H_v$ in terms of the instantaneous eigenstates of the Hamiltonian $H_{el}(0)$. That is, we can write
\begin{align}
    c_{j\sigma}=\sum_n[u_{n\sigma}(j)c_n+v_{n\sigma}(j)c_n^\dagger]\,,
\end{align}
such that $H_{el}(v=0)=\sum_{n}\epsilon_n(v=0)(c_n^\dagger c_n-1/2)$, and 
\begin{align}
    &H_v=-i\frac{\hbar v}{2}\sum_{j,\sigma}\sum_{n,m}[\partial_ju^*_{n\sigma}(j)c^\dagger_n+\partial_jv^*_{n\sigma}(j)c_n]\notag\\
    &\times[u_{m\sigma}(j)c_m+v_{m\sigma}(j)c_m^\dagger]-{\rm h. c.}\equiv-i\frac{\hbar}{2}\sum_j{\bs c}^\dagger\,R^\dagger\partial_t R\,{\bs c}\nonumber\,,
\end{align}
where ${\bs c}=(c_1,c_2,\dots, c_1^\dagger,c_2^\dagger,\dots)$ 
and $R(t)$ is the matrix whose columns are eigenvectors of ${H}_{el}(0)$. Furthermore, we have defined $\partial_ju_{n\sigma}\equiv (u_{n\sigma}(j+1)-u_{n\sigma}(j))/a$ (and similarly for $\partial_jv_{n\sigma}(j)$), while the time derivative is to be understood as $\partial_tR\equiv v\sum_j\partial_jR$.

\begin{figure}[t] 
\includegraphics[width=0.99\linewidth]{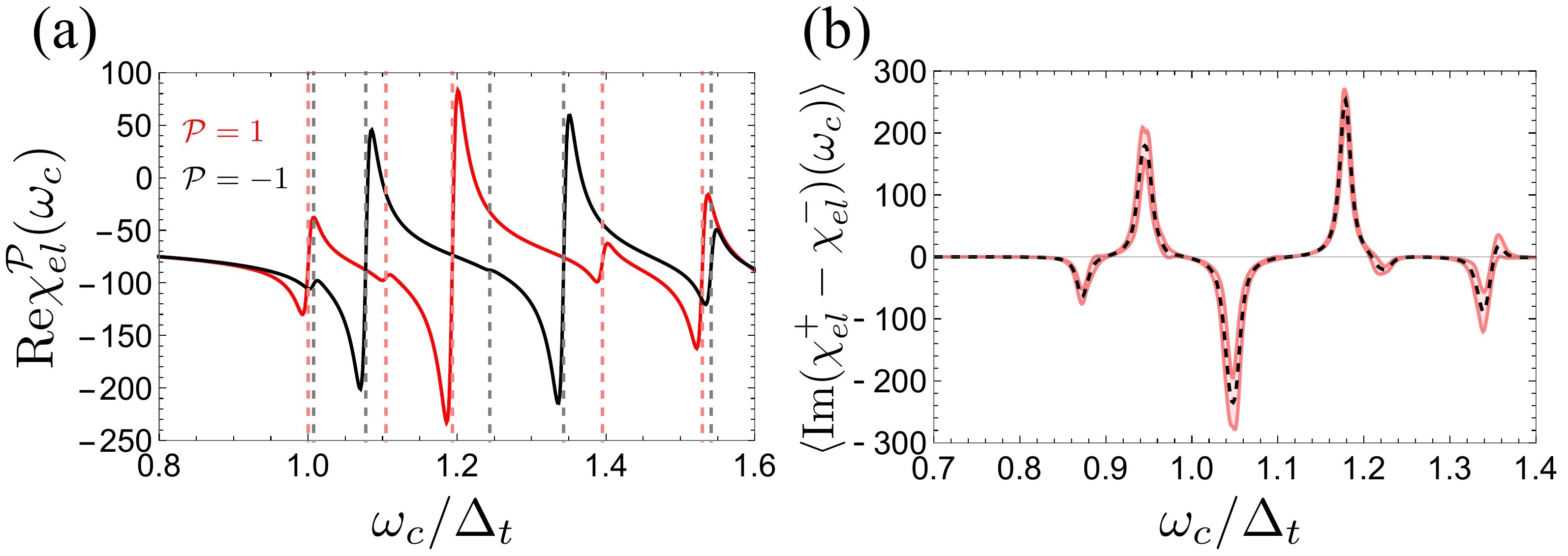}
\caption{(a) Real part of the total susceptibility ${\rm Re}\chi_{el}^\mathcal{P}(\omega_c)$ as a function of the cavity frequency $\omega_c/\Delta_t$ in topological phase ($V_Z/\Delta = 2.5$) for a finite-length nanowire homogeneously coupled to a microwave cavity. The red (black) line corresponds to the even (odd) parity of the MZMs. Pink (gray) dashed lines signal the bulk-Majorana transitions for even (odd) parity corresponding only to non-zero matrix elements. (b) Difference between the odd and even parity susceptibilities ${\rm Im }(\chi_{el}^+-\chi_{el}^-)$ of a nanowire with disorder in the chemical potential (standard deviation $\sigma_\mu = 0.1\Delta$) as a function of $\omega_c/\Delta_t$ in the topological phase ($V_Z/\Delta = 2.5$). The black dashed line corresponds to the mean value of the difference ${\rm Im}(\chi_{el}^+-\chi_{el}^-)$ calculated for $12$ disorder configurations. The pink area gives the $95 \%$ confidence interval for the population interval estimated for the difference between the two parities. The other parameters are fixed as $N=100$, $\Delta/t_h = 0.15$, $V_Z=2.5\Delta$, $\alpha/t_h = 0.4$, $\mu = 0$, and $\eta/t_h = 10^{-3}$. } 
\label{fig:Fig2}
\end{figure}

In Fig.~\ref{fig:Fig2}(a) we plot the real part of the susceptibility ${\rm Re}\chi_{el}^\mathcal{P}(\omega_c)$ for a finite-length nanowire with $N = 100$ lattice sites. We note that ${\rm Re}\chi_{el}^\mathcal{P}(\omega_c)$ exhibits different oscillation patterns for the two parities, similarly to the imaginary part of the susceptibility ${\rm Im}\chi_{el}^\mathcal{P}(\omega_c)$ described in the main text. This contribution ${\rm Re}\chi_{el}^\mathcal{P}(\omega_c)$ alters the cavity resonance frequency. To further demonstrate that the susceptibility remains a good quantity to distinguish between two parities even in the presence of disorder, we evaluate the difference $r_\chi\equiv\langle{\rm Im}(\chi_{el}^\mathcal{+}-\chi_{el}^\mathcal{-})\rangle$ for a disordered nanowire averaged over 12 disorder realizations of the chemical potential.  In  Fig.~\ref{fig:Fig2}(b) we show the disorder averaged $r_\chi$ as a function of the cavity frequency $\omega_c$. We note that the susceptibility for two parities has different intensity when the on-site fluctuations in $\mu$ are smaller than the topological gap $\Delta_t$. Therefore, the susceptibility could be utilized to distinguish between the even and the odd parities even when we average over various disorder realizations.

\begin{figure}[t] 
\includegraphics[width=0.99\linewidth]{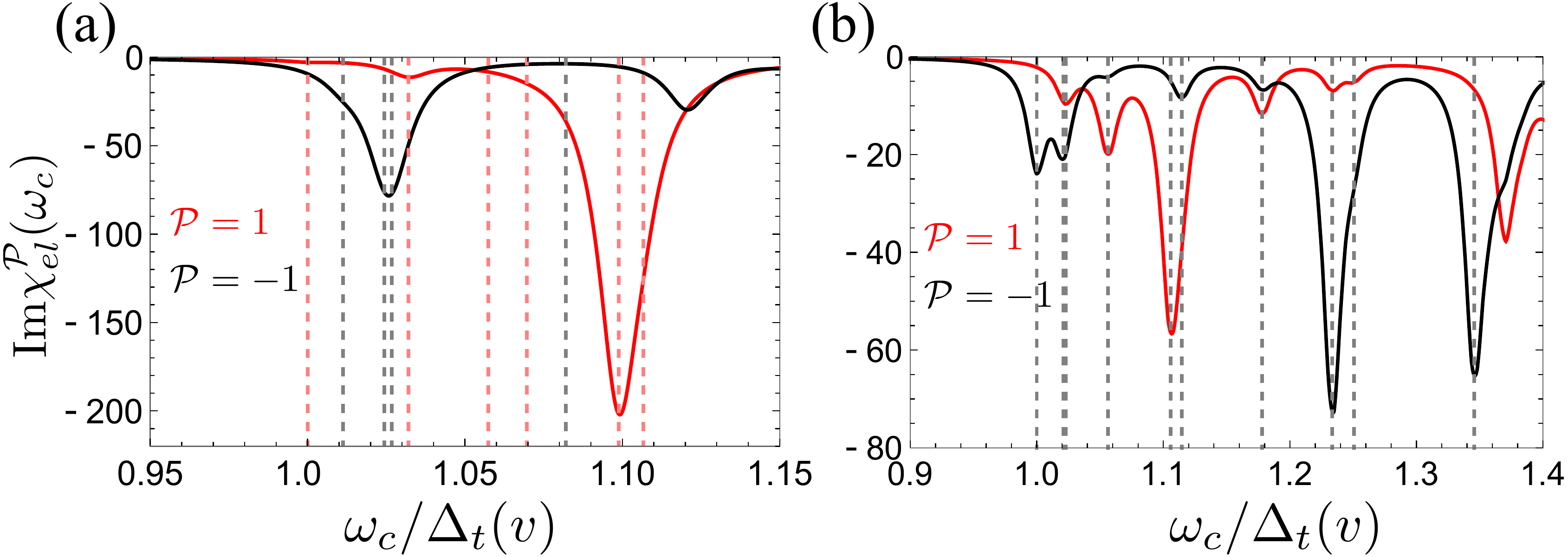}
\caption{Imaginary part of the total susceptibility ${\rm Im}\chi_{el}^\mathcal{P}(\omega_c)$ of the superconducting ring as a function of the cavity frequency normalized by the velocity-dependent topological gap $\omega_c/\Delta_{t}(v)$ for (a) $v = 0$ [$\Delta_t(v)/\Delta_t = 0.86$] and (b) $v=0.025$ [$\Delta_t(v)/\Delta_t = 0.74$]. The red (black) line corresponds to the even (odd) parity of the MZMs. The pink (gray) dashed lines indicate the positions of the peaks in ${\rm Im}\chi^\mathcal{P}_{el}(\omega_c)$ for even (odd) parity. The other parameters are fixed as $V_Z/\Delta = 2.5$, $N_{top}=100$, $N_{total}=300$, $\Delta/t_h = 0.15$, $\bar\alpha/t_h = 0.4$, $\mu/t_h = 1.5$, and $\eta/t_h = 10^{-3}$. }
\label{fig:Fig4}
\end{figure}

Additionally, in Fig.~\ref{fig:Fig4} we show ${\rm Im}\chi_{el}^\mathcal{P}(\omega_c)$ for a nanowire in a ring geometry for two different values of velocity $v$ compared to Fig.~\ref{fig:DynamicsRotated} in the main text. We find that the number and position of the peaks in ${\rm Im}\chi_{el}^\mathcal{P}(\omega_c)$ are modified for finite values of  $v$ due to the modification of the nanowire bulk wave-functions.

\section{Analytical model in the strong SOI regime} \label{AppendixC}

Here we provide more details on the derivation of the susceptibility in the continuum model. The electronic fields $\Psi_l(x)=(\psi_{l,u}(x), \psi_{l,d}(x))$ associated with the low energy  BdG Hamiltonian
\begin{align}
    \tilde{H}^l_{BdG}(x)=u p\Sigma_z+\Delta_{t}(x,0)\Sigma_x-vp\,,
    \label{ham_SM}
\end{align}
are
\begin{align}
    \psi_{l,u}(x)&=-a_+(\psi_\uparrow(x)+\psi^\dagger_\uparrow(x))-a_-(\psi_\downarrow(x)+\psi^\dagger_\downarrow(x))\,,\\
    \psi_{l,d}(x)&=-a_-(\psi_\uparrow(x)-\psi^\dagger_\uparrow(x))-a_+(\psi_\downarrow(x)-\psi^\dagger_\downarrow(x))\nonumber\,,
\end{align}
and $\psi^\dagger_{l,u(d)}(x)=+(-)\psi^\dagger_{l,u(d)}(x)$. Therefore,  the second-quantized interaction Hamiltonian between the cavity and the electronic density becomes:
\begin{equation}
    \mathcal{H}^l_{e-p}\approx \frac{2g\mu_0}{B}\int dx\psi_{l,u}(x)\psi_{l,d}(x)(a^\dagger+a)\,.
\end{equation}
We can express these fermionic fields in terms of the eigen-bogoliubons for a given chemical potential landscape
\begin{align}
    \psi_{l,u}(x)&=\sum_n[U_{n}(x)c_n+U^*_{n}(x)c_n^\dagger]\,,\\
\psi_{l,d}(x)&=\sum_n[V_{n}(x)c_n-V^*_{n}(x)c_n^\dagger]\,,
\end{align}
where $U_{n}(x)$ and $V_{n}(x)$ are the electronic weights of the state $n$ associated to the eigen-bogoliubons $c_n$ that  diagonalise the nanowire Hamiltonian, and which can in turn be written as:
\begin{align}
    c_n&=\int dx[U_n^*(x)\psi_{l,u}(x)+V_n^*(x)\psi_{l,d}(x)]\,.
\end{align}
For the MZM instead, we can write it as $c_M=(\gamma_L+i\gamma_R)/\sqrt{2}$, with $\gamma_{L,R}^\dagger=\gamma_{L,R}$ being the corresponding left and right Majorana fermion operators, with the explicit expressions
\begin{align}
    \gamma_{L}&=\int dx[\underbrace{(U_M^*(x)+U_M(x))}_{U_{L}(x)}\psi_{l,u}(x)\notag\\
    &+\underbrace{(V_M^*(x)-V_M(x))}_{iV_L(x)}\psi_{l,d}(x)]\,,\\
    \gamma_{R}&=\int dx[\underbrace{(U_M^*(x)-U_M(x))/i}_{U_{R}(x)}\psi_{l,u}(x)\notag\\
    &+\underbrace{(V_M^*(x)+V_M(x))/i}_{iV_R(x)}\psi_{l,d}(x)]\,.
\end{align}
To evaluate the susceptibility, we will employ several approximations. First, we only retain the cross terms involving MZMs and bulk states, and neglect the purely bulk contribution, focusing at frequencies $\omega<2\Delta_{t}$. Moreover, we assume the nanowire to be long enough such that the MZM splitting $\epsilon_M\approx0$. Then, the susceptibility takes the form in Eq.~\ref{susc_append}, with:  
\begin{align}
    &\mathcal{M}_{nM}^{e}=-\left(\frac{2g\mu_0}{B}\right)^2\int dx[U^*_{M}(x)V^*_n(x)-V^*_{M}(x)U^*_n(x)]\,,\\
    &\mathcal{M}_{nM}^{o}=-\left(\frac{2g\mu_0}{B}\right)^2\int dx[U_{M}(x)V^*_n(x)+V_{M}(x)U^*_n(x)]\nonumber\,.
    \label{mat_elem_low_en}
\end{align}

To determine the electron and hole weights in the above expressions, we consider the following topological landscape (in the moving frame):
\begin{align}
\Delta_{t}(x)&=\left\{
\begin{array}{cc}
\Delta_{t}\,, & x\in(-\infty,0){\rm U}(L,\infty)\\\\
-\Delta_{t}\,, & x\in[0,L]\,,
\end{array}
\right.
\end{align}
with $L$ the size of the middle part of the nanowire.  Therefore, in a given region the wave-function pertaining to the Hamiltonian in Eq.~\eqref{ham_SM} (traveling or evanescent) can be written as
\begin{align}
\psi_{\pm}(E,x)=\left(
    \begin{array}{c}
U_{\pm}(E)\\
V_{\pm}(E)
    \end{array}
\right)e^{ip_{\pm}(E)x}\,,    
\end{align}
where
\begin{align}
    p_{\pm}(E)&=\frac{\beta E\pm\sqrt{E^2-1}}{u\sqrt{1-\beta^2}}\,,\\
        V_{\pm}(E)&={\rm sign}(\Delta_{t})C(\beta)(E\mp\sqrt{E^2-1})U_{\pm}(E)\,,
\end{align}
with $\beta=v/u$ and $C(\beta)=\sqrt{\frac{1-\beta}{1+\beta}}$ being a contraction factor. Note that the effective gap is $\Delta_{t}(v)=|\Delta_{t}|\sqrt{1-\beta^2}$, and all energies above  are expressed in terms of $\Delta_{t}(v)$. However, the boundaries will lead to mixing of these pristine states with a given energy. In the following, we evaluate the degree of mixing using a scattering approach, both for the in-gap states as well as for the traveling modes. 
\subsection{In-gap Majorana modes}
To find the  spectrum and wave-functions of the in-gap modes, we set  $|E|<\Delta_{t}(v)$. Then 
\begin{equation}
    V_{\pm}(E)={\rm sign}(\Delta_{t})C(\beta)(E\mp i\sqrt{1-E^2})U_{\pm}(E)\,.
\end{equation}
with the two (complex) momenta at energy $E$:
\begin{align}
    p_{\pm}(E)=k_v(E)\pm i\kappa(E)\,,
\end{align}
where $k_v(E)=\beta E/(u\sqrt{1-\beta^2})$ and $\kappa_v(E)=\sqrt{1-E^2}/(u\sqrt{1-\beta^2})$.  Consequently, the wave-function (keeping only the terms that do not diverge at $\pm\infty$) can be written as
\begin{widetext}
\begin{align}
\psi_M(x)&=\left\{
\begin{array}{cc}
a_M \left(1, C(\beta)(E+i\sqrt{1-E^2})\right)^T e^{ip_- x}\,, & x<0\\\\
b_M \left(1, -C(\beta)(E+i\sqrt{1-E^2})\right)^T e^{ip_- x}+c_M \left(1, -C(\beta)(E-i\sqrt{1-E^2})\right)^T e^{ip_+ x}\,, & x\in[0,L]\,,\\\\
d_M \left(1, C(\beta)(E-i\sqrt{1-E^2})\right)^Te^{ip_+(x-L)}\,, & x>L\,,
\end{array}
\right.
\end{align}
\end{widetext}
from where the in-gap spectrum can be found by imposing the continuity of the wave-function at $x=0$ and $x=L$ and  that the determinant formed by the coefficients vanishes. We obtain:
\begin{equation}
    \epsilon_{M}=\displaystyle{e}^{-\frac{\sqrt{1-\epsilon_M^2}}{u\sqrt{1-\beta^2}}L}\,,
\end{equation}
and 
\begin{align}
    &b_M=\frac{ia_M\epsilon_M}{\sqrt{1-\epsilon_M^2}},\\
    &c_M=a_M\left(1-\frac{i\epsilon_M}{\sqrt{1-\epsilon_M^2}}\right),\\
    &d_M=a_M\left(\epsilon_M+i\,\sqrt{1-\epsilon_M^2}\right)\,.
\end{align}
Finally, $a_M$ can be found from the normalization condition of the in-gap states $\int_{-\infty}^\infty dx|\psi_M(x)|^2=1$ which gives:
\begin{equation}
    |a_M|\approx\frac{1}{2}\sqrt{\frac{C(\beta)}{u}}\,.
\end{equation}

\subsection{Extended bulk states}

In this case $|E|>\Delta_{t}(\beta)$, and we have to account for both the left ($+$) and right ($-$) moving states,
\begin{equation}
    V_{\pm}(E)={\rm sign}(\Delta_{t})C(\beta)(E\mp\sqrt{E^2-1})U_{\pm}(E)\,,
\end{equation}
and
\begin{align}
    p_{\pm}(E)&=\frac{\beta E\pm\sqrt{E^2-1}}{u\sqrt{1-\beta^2}}\,,
\end{align}
while the normalization of the states gives:
\begin{align}
    U_{\pm}(E)=\left(1+C^2(\beta)(2E^2-1\mp2E\sqrt{E^2-1})\right)^{-1/2}\,.
\end{align}
The positive energy solutions  for the states traveling from left to right read
\begin{widetext}
\begin{align}
\Psi_+(E,x)&=\left\{
\begin{array}{cc}
U_+(E)\left(1, C(\beta)(E-\sqrt{E^2-1})\right)^T e^{ip_+x}+a^+_E\left(1, C(\beta)(E+\sqrt{E^2-1})\right)^T e^{ip_-x}\,, & x<0\\\\
b^+_E\left(1, -C(\beta)(E-\sqrt{E^2-1})\right)^T e^{ip_+ x}+c_E^+\left(1, -C(\beta)(E+\sqrt{E^2-1})\right)^T e^{ip_- x}\,, & x\in[0,L]\,,\\\\
d_E^+\left(1, C(\beta)(E-\sqrt{E^2-1})\right)^Te^{ip_+x}\,, & x>L\,,
\end{array} 
\right.
\end{align}
and similarly for the states impinging on the topological region from the left 
\begin{align}
\Psi_-(E,x)&=\left\{
\begin{array}{cc}
a_E^-\left(1, C(\beta)(E+\sqrt{E^2-1})\right)^T e^{ip_-x}\,, & x<0\\\\
b_E^-\left(1, -C(\beta)(E-\sqrt{E^2-1})\right)^T e^{ip_+ x}+c_E^- \left(1, -C(\beta)(E+\sqrt{E^2-1})\right)^T e^{ip_- x}\,, & x\in[0,L]\,,\\\\
U_-(E)\left(1, C(\beta)(E+\sqrt{E^2-1})\right)^Te^{ip_-x}+d_E^-\left(1, C(\beta)(E-\sqrt{E^2-1})\right)^Te^{ip_+x}\,, & x>L\,.
\end{array} 
\right.
\end{align}
\end{widetext}
By imposing the continuity of the wave-functions at $x=0,L$, we can determine all the scattering coefficients $a_E^{\pm}$, $b_E^{\pm}$, $c_E^{\pm}$ and  $d_E^{\pm}$. Inserting the above wave-functions in  Eq.~D9  allows us to calculate fully analytically the susceptibility, and extract the expressions showed in the main text.

\end{document}